\documentclass[11pt,a4paper]{article}
\usepackage{jheppub,bm,booktabs,multirow}

\usepackage{color}
\usepackage{overpic}
\allowdisplaybreaks

\makeatletter
\def\@fpheader{~}
\makeatother

\def\e{\epsilon}
\def\nno{\nonumber}

\title{Non-global and rapidity logarithms in narrow jet broadening}
\author[a]{Thomas Becher,}
\author[a]{Rudi Rahn}
\author[a]{and Ding Yu Shao}
\affiliation[a]{Albert Einstein Center for Fundamental Physics, Institut f\"ur Theoretische Physik, Universit\"at Bern,
  Sidlerstrasse 5, CH-3012 Bern, Switzerland}

\emailAdd{becher@itp.unibe.ch}
\emailAdd{rahn@itp.unibe.ch}
\emailAdd{shao@itp.unibe.ch}

\date{\today}

\abstract{We derive an all-order factorization theorem for the narrow jet broadening event shape, a measure of the transverse momentum in jet events. This is a non-global observable which receives logarithmically enhanced contributions associated with the large rapidity difference between soft and collinear radiation and which is also sensitive to soft recoil effects. Our work is the first factorization analysis of an observable of this type and we show that with regard to the non-global nature, the rapidity logarithms do not constitute an essential complication since they can be tied to the jet function, which is the same as for global observables. As a consequence, the leading non-global logarithms in narrow jet broadening are encoded in the same overall factor relevant for the hemisphere soft function and light jet mass.}

\begin{document}

\maketitle

\section{Introduction}

Event shape variables are inclusive observables which measure simple geometric properties of collider events. Many of the classic  $e^+e^-$-collider event shapes are defined using the thrust axis $\vec{n}_T$ found by maximizing the thrust $T= \sum_i | \vec{n}_T \cdot \vec{p}_i|/Q$, where the sum runs over all particles in an event, and $Q$ is the center-of-mass energy. An example is the total jet broadening $B_T = \sum_i | \vec{n}_T \times \vec{p}_i|/(2Q)$ which measures the momentum transverse to $\vec{n}_T$.
Event shapes are perturbatively calculable, but higher-order terms are enhanced by large logarithms for two-jet configurations with small invariant masses. It is well known how to resum these logarithms using the factorization properties of QCD amplitudes in the soft and collinear limits. For thrust, the relevant factorization theorem was obtained in \cite{Berger:2003iw,Fleming:2007qr,Schwartz:2007ib,Bauer:2008dt} and the thrust distribution was resummed up to N$^3$LL accuracy in \cite{Becher:2008cf,Abbate:2010xh} using Soft-Collinear Effective Theory (SCET) \cite{Bauer:2000yr,Bauer:2001yt,Beneke:2002ph} (see \cite{Becher:2014oda} for a review). The factorization formula for total broadening \cite{Chiu:2011qc,Becher:2011pf,Chiu:2012ir} is more involved because for broadening the transverse momentum and virtuality of the soft radiation is comparable to the one of the energetic collinear particles. The effective theory for this kinematical situation is called SCET$_{\rm II}$ to distinguish it from the one relevant for thrust called SCET$_{\rm I}$. Because of its comparable transverse momentum, collinear radiation recoils against soft partons in SCET$_{\rm II}$. Furthermore, one encounters rapidity logarithms which are not captured using standard RG evolution from higher to lower virtuality. Formalisms to deal with this complication are available \cite{Becher:2010tm,GarciaEchevarria:2011rb,Chiu:2011qc,Collins:2011zzd,Chiu:2012ir,Vladimirov:2017ksc} and the resummation for $B_T$ has been performed at NNLL accuracy \cite{Becher:2012qc}. 

The thrust axis $\vec{n}_T$ splits the final state into two hemispheres, which we label as ``left" and ``right" and it is interesting to define separate event shapes for the partons in the two hemispheres. The left (right) broadening is defined as the sum of the absolute values of the transverse momenta of partons with the thrust axis $\vec{n}_T$ in the left (right) hemisphere 
\begin{align}
b_{L(R)} = \frac{1}{2} \sum_{i \in L(R)} |  \vec p_i^{\, \perp} | = \frac{1}{2}  \sum_{i \in L(R)}  | \vec p_i \times \vec n_T |\,.
\end{align}
For our effective theory analysis, it is natural to work with the dimensionful quantities $b_{L(R)}$, the associated dimensionless rations will be denoted by capital letters, $B_{L(R)}=b_{L(R)}/Q$. Three different combinations of left and right broadenings were measured experimentally. They are the
\begin{align*}
\text{total broadening:}\;\; b_T &=  b_L + b_R \, ,\\
\text{wide broadening:}\;\; b_W &=  {\rm max}(b_L,b_R) \, ,\\
\text{narrow broadening:}\;\; b_N &=  {\rm min}(b_L,b_R)  \,.
\end{align*}
Similarly, one can look at the invariant masses $M_L$ and $M_R$ of the jets in the two hemispheres and define the total, heavy, and light jet masses, which are the equivalent of the three quantities introduced for the broadening. Due to the left-right symmetry, the narrow broadening can be inferred from the left broadening after subtracting the wide  broadening
\begin{align}\label{eq:leftToNarrow}
\frac{d\sigma}{d b_N} = 2 \frac{d\sigma}{d b_L} - \frac{d\sigma}{d b_W}\Bigg |_{b_L=b_W=b_N}.
\end{align}
Below we analyze the factorization theorem for left broadening in the limit $b_L \ll b_R \sim Q$. 

While the heavy jet mass and the wide broadening fulfill factorization theorems analogous to the ones for total broadening and thrust, it turns out that the structure of logarithms for the left jet mass and the left broadening are much more complicated. These observables are non-global since they are only sensitive to radiation in the left hemisphere, and this induces an intricate pattern of logarithms which was discovered by Dasgupta and Salam in an analysis of the left jet mass \cite{Dasgupta:2001sh}. These authors were also able to resum the leading non-global logarithms in the large $N_c$ limit. More recently, we have derived all-order factorization formulas for non-global observables using SCET \cite{Becher:2015hka,Becher:2016mmh,Becher:2016omr}. In particular, we have analyzed the case of the left jet mass in detail in \cite{Becher:2016omr} and have shown that this observable factorizes into hard functions $\bm{\mathcal{H}}_{m}$ describing $m$ hard partons in the right hemisphere times soft functions $\bm{\mathcal{S}}_{m}$, which are given by Wilson lines along the hard partons. The complicated pattern of logarithms arises because even at leading-logarithmic accuracy, one needs to include contributions from operators with arbitrarily high multiplicity $m$.  

The non-global observables considered before are all in the SCET$_{\rm I}$ category and it is interesting to extend the results to the SCET$_{\rm II}$ case. To do so, we analyze the narrow broadening in the present paper. The relevant factorization theorem will be presented in Section \ref{sec:fact} and we confirm it by explicit computations to NNLO in Section \ref{sec:ingredients}. We resum the narrow broadening to NLL in Section \ref{sec:resum}, compare to  experimental measurements from LEP and derive the leading nonperturbative corrections affecting the distribution.


\section{Factorization formula\label{sec:fact}}

In SCET$_{\rm II}$ jet and soft functions are not well defined without an additional rapidity regulator. The divergences in this regulator cancel among the jet and soft functions leaving behind rapidity logarithms. With the regulator in place, we can derive a factorization formula following the same steps as we did for the light-jet mass and cone-jet cross sections. The key observation is that the hard partons in the unobserved right hemisphere can emit soft partons into the left hemisphere. These emissions are described by soft Wilson lines along the hard partons so that we end up with the factorization formula
\begin{align}\label{eq:fact}
\frac{d\sigma}{d b_L} = \sum_{f=q,\bar q, g} \int d b_L^s \int d^{d-2}p_L^\perp \, \mathcal{J}_f(b_L - b_L^s, p_L^\perp) \sum_{m=1}^\infty \langle \bm{\mathcal{H}}^f_m(\{\underline{n}\},Q)  \otimes\bm{\mathcal{S}}_{m}(\{\underline{n}\},b_L^s, - p_L^\perp) \rangle\,,
\end{align} 
where the hard function $\bm{\mathcal{H}}^f_m(\{\underline{n}\},Q) $ describes $m$ hard partons flying along the directions $\{\underline{n}\}=\{n_1, \dots ,n_m\}$ into the right hemisphere and a single energetic parton along $\bar{n}^\mu =(1,-\vec{n}_T)$ to the left. The soft function $\bm{\mathcal{S}}_{m}$ is given by Wilson lines along these $m+1$ partons and the jet function $ \mathcal{J}_f$ describes the splitting of the left parton with flavor $f$ into a low-mass jet. The symbol $\otimes$ indicates than one has to integrate over the direction of the hard partons and $ \langle \dots \rangle$ denotes the color trace, see \cite{Becher:2016mmh} for details on the notation and a derivation of the multi-Wilson-line structure from SCET. Note that none of the factorization discussion is affected by the presence of the regulator \cite{Becher:2011dz} which is only applied to the phase-space integrals but leaves the amplitudes unchanged. However, due to the regulator the product of soft and jet functions has implicit dependence on the hard scale $Q$. This dependence, called the collinear anomaly \cite{Becher:2010tm}, will be made manifest below. 

The hard function has the same operator definition as in the light jet mass case
\begin{align}
\label{eq:LightJetHard}
\bm{\mathcal{H}}^f_{m}(\{\underline{n}\},Q) 
  = \frac{1}{2Q} \prod_{j=1}^m &\int \! \frac{dE_j \,E_j^{d-3} }{(2\pi)^{d-2}} \,  | \mathcal{M}^{f}_{m+1}(\{p_0,\underline{p}\}) \rangle \langle \mathcal{M}^{f}_{m+1}(\{p_0,\underline{p}\}) |  
  \nonumber \\
&\times    {\Theta }_{R}\!\left(\left\{\underline{p}\right\}\right)
   (2\pi)^d \,\delta(Q-E_{\rm tot}) \,\delta^{(d-1)}(\vec{p}_{\rm tot}) \,,
\end{align}
where  $p_0^\mu= Q\, \bar{n}^\mu/2$ is the momentum of the single
hard parton of flavor $f\in\{q,\bar{q},g\}$ in the left hemisphere, and the amplitudes $|
\mathcal{M}^f_{m+1}(\{p_0,\underline{p}\}) \rangle$ are standard QCD
amplitudes for the decay of a virtual photon into $(m+1)$ partons. The function ${\Theta }_{R}\!\left(\left\{\underline{p}\right\}\right)$ enforces that the $m$ partons with momenta $\left\{\underline{p}\right\}$ are in the right hemisphere.

The associated soft function has the form
\begin{align}\label{eq:SmDefLight}
   \bm{\mathcal{S}}_m(\{ \underline{n}\},b_L, &p_L^\perp) = \int\limits_{X_s,{\rm reg}}\hspace{-0.80cm} \sum\,  \delta\Big(b_L - \mbox{$\frac{1}{2}$} \sum_{i \in X_L} |p_{L,i}^\perp|\Big)\, \delta^{d-2}(p_{X_L}^\perp - p_L^\perp) \,
   \nonumber \\
  & \times
 \langle 0 | \,\bm{S}^\dag_0(\bar{n})\,\bm{S}^\dag_1(n_1) \dots {\bm S}^\dag_m(n_m)\,  | X_s \rangle  \langle X_s| \,\bm{S}_0(\bar n) \,\bm{S}_1(n_1) \dots {\bm S}_m(n_m)\, |0\rangle \,.    
\end{align}
The integrals over phase space are regularized using the regulator \cite{Becher:2011dz}, whose explicit form will be given when we compute the one-loop soft function in \eqref{eq:softOne}. This function contains two $\delta$-function constraints: the first one fixes the contribution to the left broadening and the second one the total transverse momentum. The second constraint is necessary due to recoil effects. Only the total transverse momentum in each hemisphere vanishes, so that the soft and collinear radiations carry equal an opposite transverse momentum, see \eqref{eq:fact}. We therefore need to compute the soft function for a fixed transverse momentum of the collinear radiation. The jet function $\mathcal{J}_f(b_L - b_L^s, p_L^\perp)$ is the same as the one relevant for the total broadening and its operator definition can be found in (4) of \cite{Becher:2011pf}.

To perform the resummation it is best to Laplace transform $b_L$  and Fourier transform the transverse momentum $p_L^\perp$. Using rotation invariance around the thrust axis, the dependence on the broadening and the transverse momentum then translates into two variables $\tau_L$ and $z_L$ \cite{Becher:2011pf} and in Laplace-Fourier space the factorization formula has the simple form
\begin{align}\label{LFxsec}
\frac{d\sigma}{d\tau_L} =  \sum_{f=q,\bar q, g} \int_0^\infty d z_L \,\overline{\mathcal{J}}_f(\tau_L,z_L)  \sum_{m=1}^\infty  \big \langle \bm{\mathcal{H}}^f_m(\{\underline{n}\},Q)  \otimes \overline{\bm{\mathcal{S}}}_{m}(\{\underline{n}\}, \tau_L, z_L) \big\rangle\, . 
\end{align}
The all-order form of the rapidity divergences was derived in the study of the total broadening in \cite{Becher:2011pf}. The divergences in the soft functions must cancel against the divergences of the jet function, leaving behind rapidity logarithms. Because these logarithms are fully determined by the divergences of the jet function, the collinear anomaly for narrow broadening must be the same form as the one of the total broadening. Extracting the anomaly logarithms, the fully factorized form of the cross section is given by
\begin{align}\label{eq:factfinal}
\frac{d\sigma}{d\tau_L} =  \sum_{f=q,\bar q, g} \int_0^\infty d z_L  (Q^2 \overline{\tau}_L^2)^{-F^f_B(\tau_L,z_L,\mu)} \frac{z_L}{(1+z_L^2)^{3/2}}  \sum_{m=1}^\infty  \big \langle \bm{\mathcal{H}}_m^f(\{\underline{n}\},Q)  \otimes \overline{\bm{\mathcal{W}}}_{m}^{\, f}(\{\underline{n}\}, \tau_L, z_L) \big\rangle \,,
\end{align}
where the refactorized functions $\overline{\bm{\mathcal{W}}}_{m}^{\, f}$ are independent of $Q$ and are defined by the product of the jet and soft functions
\begin{align}
\overline{\mathcal{J}}_f(\tau_L,z_L,\mu) \overline{\bm{\mathcal{S}}}_{m}(\{\underline{n}\}, \tau_L, z_L,\mu) = (Q^2 \overline{\tau}_L^2)^{-F_B^f(\tau_L,z_L,\mu)} \frac{z_L}{(1+z_L^2)^{3/2}} \overline{\bm{\mathcal{W}}}_{m}^{\,f}(\tau_L,z_L,\mu)\,.
\end{align}
We have extracted the LO jet function so that $\overline{\bm{\mathcal{W}}}_{m}^{\, f}=\bm{1}+\mathcal{O}(\alpha_s)$. The anomaly exponent for the quark case $F^q_B$ is the one encountered in the total broadening which was computed to two loops in \cite{Becher:2012qc}. The one for the gluon channel is related to it by Casimir scaling $F^g_B = C_A/C_F  \,F^q_B$ up to three-loop accuracy.

While the rapidity divergences must cancel in \eqref{LFxsec}, this does not guarantee that all the functions $\overline{\bm{\mathcal{W}}}_{m}^{\,f}$ are finite. In principle, there could be divergences in these functions which only vanish after integrating over angles and combining different multiplicities. However, in our explicit one and two-loop computations in the next section we find that the functions $\overline{\bm{\mathcal{W}}}_{m}^{\,f}$ are finite.

\section{Ingredients of the factorization theorem and collinear anomaly\label{sec:ingredients}}

It is interesting to compute the ingredients of the factorization formula perturbatively to explicitly verify the above structure. Fortunately, many of the ingredients are already known. The hard functions are the same as the ones for the light-jet-mass case \cite{Becher:2016omr} and were given in Section 4 of this paper. The jet function $\mathcal{J}_q$ is the same as the one for the total broadening calculated at one-loop order in \cite{Becher:2012qc} using the analytical phase-space regulator \cite{Becher:2011dz} to regularize the rapidity divergences. We write 
\begin{align}
\overline{\mathcal{J}}_f(\tau,z)  = \overline{\mathcal{J}}_f^{(0)}(\tau,z) \left [1 + \frac{\alpha_s}{4\pi}\, \overline{\mathcal{J}}_f^{(1)}(\tau,z) \right] ,
\end{align}
where the tree-level result is given by
\begin{align}
 \overline{\mathcal{J}}_{q}^{(0)}(\tau,z) =  \overline{\mathcal{J}}_{g}^{(0)}(\tau,z)  & =  \frac{4^\e \, \Gamma(2-2\e)}{\Gamma^2(1-\e)} \frac{z^{1-2\e}}{(1+z^2)^{3/2-\e}} \,, 
 \end{align}
 and the divergent part of the one loop result for the quark case reads
 \begin{align}
\overline{\mathcal{J}}_q^{(1)}(\tau,z) & =  C_F (\mu^2 \bar \tau^2)^\e (\nu Q \bar \tau^2)^\alpha 
  \left\{ - \frac{4}{\alpha} \Big( \frac{1}{\e} + 2 \ln z_+ \Big)  + \frac{4}{\e^2} + \frac{3}{\e} \right\}.
\end{align}
The dimensional regulator is introduced as $d=4-2\epsilon$ and $\alpha$ regularizes the rapidity divergence. The full one-loop quark jet function can be found in \cite{Becher:2012qc}. We do not need the expression for the one-loop gluon jet function because the associated hard function is suppressed by $\alpha_s$.

The only new ingredient are the soft functions $\bm{\mathcal{S}}_m$.  At the one-loop order, they are given by sums of dipole factors
\begin{align}\label{eq:softOne}
\bm{\mathcal{S}}_m =   &\,  \delta(b_L) \delta^{d-2}(p_L^\perp) \, \bm{1} -  g_s^2 \tilde\mu^{2\e}  \int [ d  k] \left(\frac{\nu}{k^+}\right)^\alpha  \left[  \sum_{i = 1}^m (\bm{T}_0 \cdot \bm{T}_i  + \bm{T}_i \cdot \bm{T}_0)\,  \frac{\bar n \cdot n_i}{ \bar n \cdot k \,  n_i\cdot k} \right.  \nno  \\ 
&+  \left. \sum_{i,j=1}^m \bm{T}_i \cdot \bm{T}_j \,  \frac{ n_i \cdot n_j}{  n_i \cdot k \,   n_j \cdot k} \right] \delta^{d-2} (k^\perp - p_L^\perp) \delta\left(b_L - \frac{|k^\perp|}{2}\right) \theta(k^+ - k^-)\,,
\end{align}
where $[d k ] = d^d k \,  \delta^+(k^2) /(2\pi)^{d-1} $ denotes the phase-space integration.  The scale $\tilde\mu^2 = \mu^2 e^{\gamma_E}/(4\pi)$ and the light-cone components are $k^+ = k\cdot n$ and $k^- = k\cdot \bar{n}$.  The factor $(k^+)^{-\alpha}$ regularizes the rapidity divergences which can only arise in the terms in the first line involving the left Wilson line with color structure $\bm{T}_0$ along the $\bar n$-direction. Let us now focus on these terms, which involve the integral 
\begin{align}
I_i = g_s^2 \tilde\mu^{2\e}  \int [ d  k]  \left(\frac{\nu}{k^+}\right)^\alpha  \frac{\bar n \cdot n_i}{ \bar n \cdot k \,  k \cdot n_i}     \delta^{d-2} (k^\perp - p_L^\perp) \delta\left(b_L - \frac{|k^\perp|}{2}\right) \theta(k^+ - k^-)\,,
\end{align}
To evaluate it, we introduce light-cone coordinates along the $n$ and $\bar{n}$ directions and rewrite 
\begin{align}
k \cdot n_i = \frac{1}{2} k^+ n_i^- + \frac{1}{2} k^- n_i^+ + k^\perp \cdot n_i^\perp\,,
\end{align}
with $ n_i^+ n_i^-= -(n_i^{\perp})^2=n_{iT}^2 \geq0$. 
We then change to light-cone coordinates and integrate over $k^\perp$ and $k^-$ to eliminate the two $\delta$-functions, which leads to
\begin{align}
I_i  = g_s^2 \tilde\mu^{2\e} \frac{n_i^-}{(2\pi)^{d-1}} \frac{1}{p_L^2} \delta\left(b_L - \frac{p_L}{2}\right) \int_{p_L}^\infty d k^+ \left( \frac{\nu}{k^+} \right)^\alpha \frac{1}{k^+ n_i^- + n_i^+ p_L^2/k^+ + \delta_i}\, ,
\end{align}
with $\delta_i = 2 p_L^\perp \cdot n_i^\perp = -2 p_L  n_i^T \cos\phi_i $, where $\phi_i$ is the angle between $p_L^\perp$ and $n_i^\perp$. Introducing a new variable $z = p_L/k^+$, the remaining integral takes the form
\begin{align}
p_L^{1-\alpha} \nu^\alpha \int_0^1 d z \frac{z^{-1+\alpha}}{z^2 n_i^+ p_L + z \delta_i  + n_i^- p_L} 
\end{align}
and its rapidity divergence can be extracted using the relation
\begin{align}
z^{-1+\alpha} = \frac{1}{\alpha} \delta(z) + \left[ \frac{1}{z}\right]_+ + \mathcal{O}(\alpha)\,.
\end{align}
Integrating over $z$ and combining the result with the color structure, we obtain
\begin{align}
\bm{\mathcal{S}}_m= &\, \delta(b_L) \delta^{d-2}(p_L^\perp) \,\bm{1} + \frac{\alpha_s}{4\pi} \sum_{i = 1}^m (\bm{T}_0 \cdot \bm{T}_i  + \bm{T}_i \cdot \bm{T}_0)   \frac{2(\mu^2 e^{\gamma_E})^\e}{\pi^{1-\e}} \frac{ \nu ^\alpha}{p_L^{2+\alpha}} \delta\left(b_L - \frac{p_L}{2}\right)   \nno \\
&  \times \Bigg[ \frac{1}{\alpha} + \frac{1}{2} \ln \omega_i + \left [\frac{\pi}{2}-\phi_i  - \arccos\left( \sqrt{\omega_i}\sin\phi_i\right) \right] \cot\phi_i \Bigg] + (\text{``$\bm{T}_i \cdot \bm{T}_j$ terms''}) \,,
\end{align}
where $\omega_i = n_i^-/(n_i^- + n_i^+ - 2 n_i^T \cos\phi_i)$. Using color conservation $\sum_{i=0}^m \bm{T}_i=0$, the $1/\alpha$ pole terms simplify to
\begin{align}
\bm{\mathcal{S}}_m = \,\delta(b_L) \delta^{d-2}(p_L^\perp)\, \bm{1} - \frac{\alpha_s}{4\pi}    \frac{4(\mu^2 e^{\gamma_E})^\e}{\pi^{1-\e}} \frac{ \nu ^\alpha}{p_L^{2+\alpha} }\delta\left(b_L - \frac{p_L}{2}\right) \left[ \frac{C_0}{\alpha} \, \bm{1} + \cdots \right] ,
\end{align}
where $C_0\, \bm{1}= \bm{T}_0^2$ is the quadratic Casimir of the relevant representation, $C_0 = C_F$ and $C_0 = C_A$ for quarks and gluons, respectively. 

Both the terms in the first and second line of \eqref{eq:softOne} involve a UV divergence, which becomes visible after performing the Laplace and Fourier transformations. For the $1/\alpha$ pole term, this leads to
\begin{align}\label{barSm}
\bm{\mathcal{\overline{S}}}_m = \,\bm{1} + \frac{\alpha_s}{4\pi}   (\mu \bar \tau_L)^{2\e}  (\nu \bar \tau_L)^\alpha \left[ \frac{4 \,C_0}{\alpha}\left( \frac{1}{\e} + 2 \ln z_+ \right) \bm{1}  + \cdots \right] .
\end{align}
We see that the one-loop $1/\alpha$ pole is the same for all the functions and cancels between jet and soft function, as required by the consistency of the factorization theorem.

Using the known results from the global broadening variables, we can analyze the divergences also at NNLO in the cross section. 
 Since we find that the rapidity divergences cancel out before integration over $z_L$ in (\ref{LFxsec}), it is convenient to analyze the integrand. Explicitly, we define the combination $C_f$ through
\begin{align}\label{cf}
\overline{\mathcal{J}}_f^{(0)} C_f(\tau_L,z_L) = \overline{\mathcal{J}}_f(\tau_L,z_L)  \sum_{m=1}^\infty  \big \langle \bm{\mathcal{H}}^f_m(\{\underline{n}\},Q)  \otimes \overline{\bm{\mathcal{S}}}_{m}(\{\underline{n}\}, \tau_L, z_L) \big\rangle \,,
\end{align}
factoring out the leading order jet function for convenience. 
We write the perturbative expansion of the ingredients of $C_f$ in the bare coupling constant $\alpha_0$ as
\begin{align}
& \bm{\mathcal{H}}^f_m  = \sum_{n=m-1}^\infty\left( \frac{\alpha_0}{4\pi} \right)^n \bm{\mathcal{H}}_m^{f,(n)}\,, ~~~~~
\overline{\bm{\mathcal{S}}}_m  =  \bm{1} + \sum_{n=1}^\infty \left(\frac{\alpha_0}{4\pi}\right)^n \overline{\bm{\mathcal{S}}}_m^{(n)}\,, \nno \\
& \overline{\mathcal{J}}_f  =  \overline{\mathcal{J}}_f^{(0)}\left[ 1 + \sum_{n=1}^\infty  \left(\frac{\alpha_0}{4\pi}\right)^n \overline{\mathcal{J}}_f^{(n)}  \right],
\end{align}
Substituting the above expanded expressions into (\ref{cf}), we obtain the one- and two-loop  coefficients as
\begin{align} 
C_f^{(1)} = &  \sum_{m=1}^2 \bm{\mathcal{H}}_m^{f,(1)} \otimes \bm{1} + \bm{\mathcal{H}}_1^{f,(0)} \otimes  \left[ \overline{\mathcal{J}}_f^{(1)} \cdot\bm{1}  +  \bm{\mathcal{S}}_1^{(1)}  \right],  \nno \\
C_f^{(2)} = & \sum_{m=1}^3 \bm{\mathcal{H}}_m^{f,(2)} \otimes \bm{1} + \bm{\mathcal{H}}_1^{f,(1)} \otimes  \left[ \overline{\mathcal{J}}^{(1)}_f \cdot\bm{1}  +  \bm{\mathcal{S}}_1^{(1)}  \right] + \bm{\mathcal{H}}_2^{f,(1)} \otimes \left[ \overline{\mathcal{J}}_f^{(1)} \cdot \bm{1}  +  \bm{\mathcal{S}}_2^{(1)}  \right] \nno \\
& + \bm{\mathcal{H}}_1^{f,(0)} \otimes \left[  \overline{\mathcal{J}}_f^{(2)} \cdot \bm{1} + \overline{\mathcal{J}}_f^{(1)}  \cdot \bm{\mathcal{S}}_1^{(1)} + \bm{\mathcal{S}}_1^{(2)}   \right].
\end{align}
We have split the results into several terms. The first one at each order is the purely hard contribution which is free of rapidity divergences. The remaining terms include rapidity divergences which cancel out in each square bracket.  Specifically, for the two-loop $\bm{\mathcal{S}}_1$ only double radiation inside the left hemisphere contributes to the $\alpha$ poles, which can be 
immediately extracted from wide-broadening soft function computation in \cite{Becher:2012qc}. Therefore one can easily verify that both of $[\, \overline{\mathcal{J}}_q^{(1)} \cdot\bm{1}  +  \bm{\mathcal{S}}_1^{(1)} \,] $ and $ [\,\overline{\mathcal{J}}_q^{(2)} \cdot \bm{1} + \overline{\mathcal{J}}_q^{(1)}  \cdot \bm{\mathcal{S}}_1^{(1)} + \bm{\mathcal{S}}_1^{(2)} \, ]$ are rapidity divergence free (note that only $f=q$ contributes in these combinations at NNLO).  The $1/\alpha$ divergence of the one-loop $\bm{\mathcal{S}}_2$ in (\ref{barSm})  does not depend on the directions of the hard partons described by the hard function. Therefore the rapidity divergences 
in the $[\, \overline{\mathcal{J}}_f^{(1)} \cdot\bm{1}  +  \bm{\mathcal{S}}_2^{(1)} \,] $ term also cancel before the angular convolution.  
So we conclude that up to NNLO all rapidity divergences cancel out within our factorization formula and that the associated rapidity logarithms indeed have the required structure. We also see that the cancellation of the divergences takes place already before the angular convolutions are performed, at least to NNLO.

As a final check, let us verify that our formula \eqref{eq:fact} reproduces the NLO fixed-order result for the left broadening. At NLO the wide-broadening hemisphere always includes two partons, while there is only a single parton in the narrow broadening hemisphere. Therefore the narrow broadening vanishes at $\mathcal{O}(\alpha_s)$ and we should find that the left broadening at NLO is half of the wide broadening. For the NLO result we only need one-loop $\bm{\mathcal{S}}_1$ which is related to the left-right-broadening soft function ${\cal S}$ in \cite{Becher:2012qc} by
\begin{align}
\langle \bm{\mathcal{S}}_1(\{n\},b_L,p_L^\perp) \rangle = \int \! d \, b_R\int \!d^{d-2} p_R^\perp \,  {\cal S}(b_L,b_R,p_L^\perp,p_R^\perp)\,,
\end{align}
where we have set $n_1=n$ which the hard function $\bm{\mathcal{H}}^q_1$ will enforce due to momentum conservation.
The integral over the right hemisphere simply sets the right-side contribution to zero because this part becomes scaleless.  Using the results in \cite{Becher:2012qc} we obtain 
\begin{align}
\langle \overline{\bm{\mathcal{S}}}_1(\{n\},\tau_L,z_L) \rangle = &~ 1 + \frac{\alpha_s C_F}{4\pi} (\mu^2 \bar \tau_L^2)^\e (\nu \bar \tau_L)^\alpha \Bigg[ \frac{4}{\alpha}\left( \frac{1}{\e} + 2\ln z_+^L \right) - \frac{2}{\e^2} + 8 \, {\rm Li}_2\! \left( - \frac{z_-^L}{z_+^L} \right) \nno \\
&+ 4 \ln^2 z_+^L + \frac{5\pi^2}{6} \Bigg]\,.
\end{align}
Combining all bare one-loop ingredients we reproduce the correct left-broadening distribution
\begin{align}
\frac{B_L}{\sigma_0} \frac{d\sigma}{dB_L} = \frac{\alpha_s}{4\pi}C_F \left(-8 \ln B_L -6\right),
\end{align} 
which is exactly half of the result for wide broadening. After these consistency checks, we now turn to resummation.

\section{NLL resummation\label{sec:resum}}

The resummation to NLL proceeds as in the light-jet-mass case discussed in \cite{Becher:2016omr}. The main simplification at NLL is that we only need to include the hard function $\bm{\mathcal{H}}^q_1$ for a high value of the renormalization scale $\mu \sim Q$ since the higher-multiplicity hard functions are suppressed by powers of $\alpha_s$ and do not suffer from large logarithms with this scale choice. We can then evolve the hard function $\bm{\mathcal{H}}_1^q$ from the high scale $\mu_h \sim Q$ to a low scale $\mu \sim 1/\tau_L$ to resum the logarithms and combine it with the tree-level soft and jet functions in $\overline{\bm{\mathcal{W}}}_{m}^{\,q}=\bm{1}$ and the one-loop anomaly coefficient $F_B^q$. In this approximation, the factorization formula \eqref{eq:factfinal} simplifies to 
\begin{align}
\frac{1}{\sigma_0}\frac{d\sigma}{d\tau_L} =    (\mu \bar \tau_L)^{-\eta_L} I(\eta_L) \sum_{m=1}^{\infty}  \big \langle   \bm{U}_{1m}^H(\{\underline{n}\},\mu,\mu_h,Q)\hat{\otimes} \bm{1}\big\rangle \,,
\end{align}
with $\eta_L = C_F \alpha_s(\mu) \ln(Q^2 \bar \tau_L^2)/\pi\,$. The function 
\begin{align}
I(\eta) = \frac{4^\eta}{1+\eta}\, {}_2 F_1(\eta,1+\eta,2+\eta,-1)
\end{align}
is obtained by carrying out the $z_L$ integration and  the evolution matrix takes the form
\begin{align}\label{Ufac}
\bm{U}^H(\{\underline{n}\},\mu,\mu_h,Q) = \bm{P} \exp\left[ \int_{\mu}^{\mu_h} d \ln \nu \, \bm{\Gamma}^H (\{\underline{n}\},\nu,Q) \right]. 
\end{align}
The quantity $ \bm{\Gamma}^H$ determines the RG evolution behavior of the hard functions, which are the same as in the light-jet-mass case.  In \cite{Becher:2016omr}, we split this hard anomalous dimension into two pieces
\begin{align}
\bm{\Gamma}_{lm}^H = \hat{\bm{\Gamma}}_{lm} -\left[ 2\, \Gamma_{\rm cusp}(\alpha_s) \ln\frac{Q}{\mu} + 2\gamma^J(\alpha_s) \right] \delta_{lm}\,.
\end{align}
The first part $\hat{\bm{\Gamma}}_{lm}$ induces the single-logarithmic non-global structure, while the all the double logarithms are obtained from the diagonal remainder. As a consequence, the total evolution factor \eqref{Ufac} is given by a product of a global and non-global evolution factor
\begin{align}
\sum_{m=1}^{\infty}   \big \langle   \bm{U}_{1m}^H(\{\underline{n}\},\mu,\mu_h,Q)\hat{\otimes} \bm{1}\big\rangle = U^H(\mu,\mu_h,Q) S_{\rm NG}(\mu,\mu_h)\,.
\end{align}
Up to NLL accuracy, the global evolution factor $U^H$ is the same as the evolution factor which drives the vector-current Wilson coefficient $C_V(Q^2,\mu)$ in SCET, see e.g.\ \cite{Becher:2006mr}. Explicitly, it is given by 
\begin{align}
\ln U^H(\mu,\mu_h,Q) &= 2S(\mu_h,\mu) - 2 A_{\gamma^J}(\mu_h,\mu) -  A_{\Gamma_{\rm cusp}}(\mu_h,\mu) \ln \frac{Q^2}{\mu_h^2}  \nno \\
&=\frac{2 C_F}{\beta_0^2} \Big[ \frac{4\pi}{\alpha_s(\mu_h)}\left(1 - \frac{1}{r} - \ln r\right) + \left(\frac{\gamma^{{\rm cusp}}_1}{\gamma^{\rm cusp}_ 0} - \frac{\beta_1}{\beta_0}\right)(1-r+\ln r) + \frac{\beta_1}{2\beta_0} \ln^2 r \nno \\
&~~~~+ \frac{3\beta_0}{2} \ln r - \beta_0 \ln r\, \ln \frac{Q^2}{\mu_h^2} \Big] \,,
\end{align}
where $r = \alpha_s(\mu)/\alpha_s(\mu_h)$. The non-global evolution factor is the same as in the light-jet-mass case and we use the parametrization \cite{Dasgupta:2001sh} 
\begin{equation}\label{SNG}
S_{\rm NG}(\mu,\mu_h)
   \approx \exp\!\left(-C_A C_F \frac{\pi^2}{3} \,u^2 \frac{1+(a u)^2}{1+(b u)^c}\right) ,  
\end{equation}
with 
\begin{equation}
u = \frac{1}{\beta_0}\ln \frac{\alpha_s(\mu)}{\alpha_s(\mu_h)}\,,
\end{equation}
where the constants $a=0.85\, C_A$ , $b=0.86\, C_A$, $c=1.33$ were determined by fitting to the result of a  parton-shower computation in the large-$N_c$ limit. The numerical result for $N_c=3$ was recently obtained in \cite{Hagiwara:2015bia}. Numerically, the corrections to the large-$N_c$ limit are small as long as the exact two-loop color factor is accounted for, as is done in \eqref{SNG}.

\begin{figure}[t!]
\begin{center}
\begin{tabular}{ccc}
\includegraphics[width=0.45\textwidth]{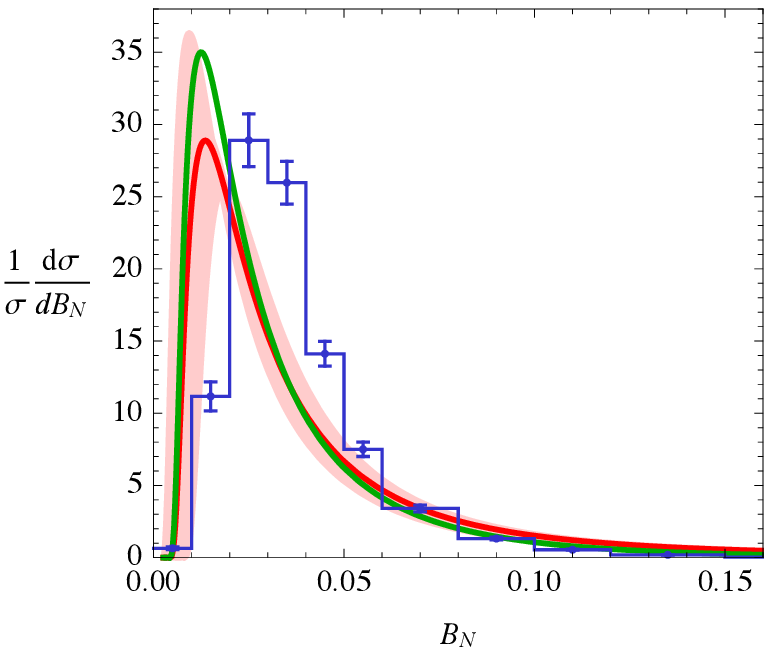} && \includegraphics[width=0.45\textwidth]{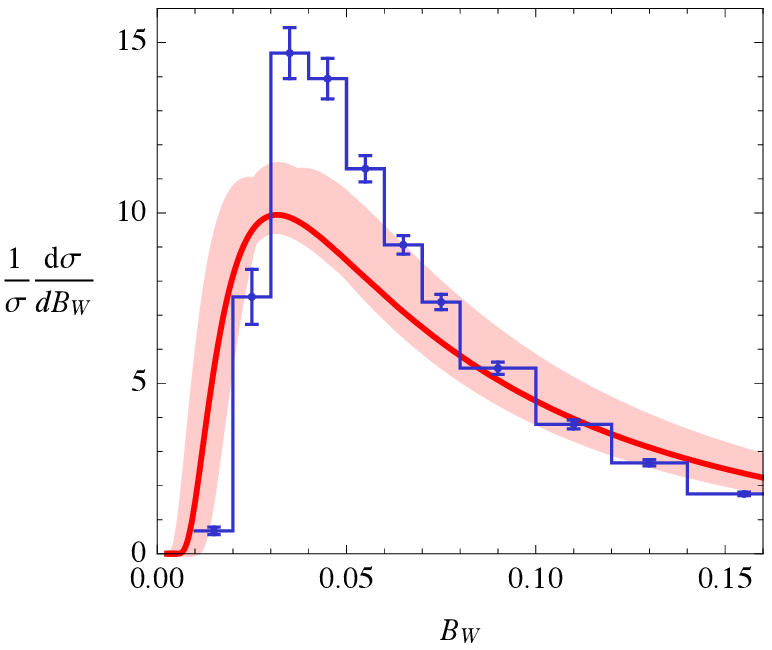} 
\end{tabular}
\end{center}
\vspace{-0.3cm}
\caption{The red bands show the NLL result for the narrow broadening (left) and the wide broadening (right), compared to {\sc Delphi} data (blue) \cite{Abreu:1996na}.  The green line is the purely global part of the narrow broadening distribution. \label{fig:NLLnarrowwide}}
\end{figure}

In the low energy range $\ln(\mu \bar \tau_L)$ counts as $\mathcal{O}(1)$ and we can approximate 
\begin{align}
\eta_L \approx \eta = \frac{C_F \alpha_s(\mu)}{\pi} \ln\frac{Q^2}{\mu^2}\,.
\end{align}
After this, we can analytically invert the Laplace transformation and obtain
\begin{align}
\frac{1}{\sigma_0}\frac{d\sigma}{d b_L} =  U^H(\mu,\mu_h,Q) S_{\rm NG}(\mu,\mu_h)
 \frac{e^{-\gamma_E \eta}}{\Gamma(\eta)} \frac{1}{b_L} \left(\frac{b_L}{\mu}\right)^{\eta} I(\eta) \,.
 \end{align}
We find that our NLL resummation formula is basically the square-root of (43) in \cite{Becher:2011pf} up to the non-global evolution factor. In order to calculate the differential distribution one can use the above equation directly or first integrate it and then take the derivative. One advantage of the latter scheme is that the resummed distribution is automatically normalized. We denote the integrated spectrum by
\begin{equation}
R(B_L) = \int_0^{Q B_L} \!\!db_L\, \frac{1}{\sigma_0}\frac{d\sigma}{d b_L}  = S_{\rm NG}(\mu,\mu_h) \,\Sigma_q(B_L) \,,
\end{equation}
where the global part is given by
\begin{equation}
\Sigma_q(B_L) = U^H(\mu,\mu_h,Q) \frac{e^{-\gamma_E \eta}}{\Gamma(\eta+1)}  \left(\frac{Q B_L}{\mu}\right)^{\eta} I(\eta) \,.
\end{equation}
As in the light-jet case, the non-global effects simply enter as a prefactor at NLL accuracy which multiplies the quantity $\Sigma_q(B_L)$ familiar from the coherent branching formalism \cite{Catani:1989ne,Catani:1990rr,Catani:1992ua}. The prefactor is absent for wide broadening, which to NLL is given by
\begin{equation}
R(B_W)= [\Sigma_q(B_W) ]^2\,.
\end{equation}
Using relation \eqref{eq:leftToNarrow}, we then obtain the narrow broadening, which can be compared to LEP measurements from the {\sc Delphi} \cite{Abreu:1996na} or {\sc OPAL} \cite{Abbiendi:2004qz} collaborations. In Figure \ref{fig:NLLnarrowwide} we show the NLL predictions, compared to the {\sc Delphi} measurements. For the plots, we use $\alpha_s(M_Z) = 0.1181$ \cite{Olive:2016xmw} and estimate the uncertainty by varying each of the scales $\mu_h$ and $\mu$ by a factor two around their default values and taking the envelope of the scale variations.

It is clear that the distributions are affected by nonperturbative effects in the peak region, and it turns out that the nonperturbative effects are logarithmically enhanced for jet broadening \cite{Dokshitzer:1998qp,Becher:2013iya}. The paper \cite{Becher:2013iya} demonstrated that the dominant effects are nonperturbative corrections to the anomaly coefficient and that these corrections are obtained from the same nonperturbative matrix element ${\cal A}$ which is responsible for the nonperturbative shift in the thrust distribution and other event shapes \cite{Lee:2006nr}. For narrow broadening, these results imply that the leading nonperturbative effects are obtained from shifting the distribution by
\begin{equation}
B_N \to B_N - \frac{{\cal A}}{2} \ln\frac{1}{B_N}
\end{equation}
and the value extracted from the thrust distribution is ${\cal A} \approx 0.3\, {\rm GeV}$ \cite{Abbate:2010xh}. Near the peak, this would imply shifts of $\Delta B_N \approx 0.007$ and $\Delta B_W \approx 0.006$ in the two distributions, in qualitative agreement with the data. 

We find it remarkable that the leading nonperturbative effects in a non-global observable
are related to the shift in thrust. The underlying mechanism is of course that the collinear anomaly connects the enhanced nonperturbative effects in the soft functions $\bm{\mathcal{S}}_m$ to the ones in the jet function, which is the same as in the global variants of broadening. Through the anomaly, this in turn is connected to the nonperturbative effect in the much simpler soft functions relevant in the global case. 

In practice, the logarithmically enhanced nonperturbative effects might not be sufficient to obtain satisfactory agreement with data, and  also non-logarithmic and non-universal shifts should be included, as well as other shape parameters. Before analyzing this further, one should include the matching to fixed-order perturbation theory and, if possible, increase the logarithmic accuracy of the resummation. We will not pursue these issues further for the moment.

\section{Conclusion}

In this short note we have analyzed the narrow broadening, a non-global, recoil-sensitive observable suffering from rapidity logarithms. We have obtained a factorization formula for this event shape which is of a similar form as the theorems for the light jet mass and the hemisphere soft function. The main result of our analysis is that the rapidity logarithms governed by the collinear anomaly are separate from the non-global structure since they can be tied to the jet function. As a consequence, the non-global effects in narrow broadening are identical to the ones for the light jet mass at NLL. We were also able to derive the leading logarithmically enhanced nonperturbative effects in narrow broadening and relate them to the nonperturbative shift in the thrust distribution. Taking these effects into account leads to reasonable agreement with LEP measurements within the limited accuracy of our calculation. 

Of course, there are other recoil-sensitive non-global observable observables which can be analyzed in our framework. An example is the jet shape introduced in \cite{Ellis:1992qq}. This case is slightly simpler in that the soft radiation only affects the observable indirectly, via the jet axis. Using a recoil-free axis \cite{Larkoski:2014uqa}  then eliminates the effect of soft radiation and the associated non-global structure \cite{Kang:2017mda}. For the narrow broadening, on the other hand, the non-global structure will persist, even with a recoil-free axis.

\begin{acknowledgments}	
The authors thank Ben Pecjak for comments on the manuscript. T.B.\ is supported by the Swiss National Science Foundation (SNF) under grant 200020\_165786, and R.R under CRSII2\_160814. The authors thank the Munich Institute for Astro- and Particle Physics (MIAPP) of the DFG cluster of excellence "Origin and Structure of the Universe" for hospitality and support.
\end{acknowledgments}

\end{document}